\documentclass[12pt]{article}
\usepackage[a4paper, total={6in, 9in}]{geometry}
\usepackage[dvipsnames]{xcolor}
\usepackage{parskip}
\usepackage{authblk}
\usepackage{graphicx} 
\usepackage[
backend=biber,
style=numeric,
sorting=nyt
]{biblatex}
\usepackage{hyperref}
\usepackage{xcolor}

\renewcommand{\footnotesize}{\scriptsize}

\title{Real-time Risk Metrics for Programmatic Stablecoin Crypto Asset-Liability Management (CALM) \footnote{The authors would like to thank Professor Moorad Choudhry for review comments on an earlier draft.}}

\author[1]{Marcel Bluhm}
\author[2]{Adrian Cachinero Vasiljevi\'c}
\author[3]{S\'ebastien Derivaux}
\author[4]{S\o ren Terp H\o rl\"uck Jessen}

\affil[1]{The Block, \href{mailto:mbluhm@theblock.co}{mbluhm@theblock.co}}
\affil[2]{Steakhouse Financial Limited, \href{mailto:adrian@steakhouse.financial}{adrian@steakhouse.financial}}
\affil[3]{Steakhouse Financial Limited, \href{mailto:sebastien@steakhouse.financial}{sebastien@steakhouse.financial}}
\affil[4]{Balloonist ApS, \href{mailto:yourfriends@balloonist.xyz}{yourfriends@balloonist.xyz}}

\date{23 January 2024}

\begin{document}
\maketitle
\begin{abstract}
     Stablecoins have turned out to be the ``killer" use case of the growing digital asset space. However, risk management frameworks, including regulatory ones, have been largely absent. In this paper, we address the critical question of measuring and managing risk in stablecoin protocols, which operate on public blockchain infrastructure. The on-chain environment makes it possible to monitor risk and automate its management via transparent smart-contracts in real-time. We propose two risk metrics covering capitalization and liquidity of stablecoin protocols. We then explore in a case-study type analysis how our risk management framework can be applied to DAI, the biggest decentralized stablecoin by market capitalisation to-date, governed by MakerDAO. Based on our findings, we recommend that the protocol explores implementing automatic capital buffer adjustments and dynamic maturity gap matching. Our analysis demonstrates the practical benefits for scalable (prudential) risk management stemming from real-time availability of high-quality, granular, tamper-resistant on-chain data in the digital asset space. We name this approach Crypto Asset-Liability Management (CALM). 

\end{abstract}
\clearpage

\section{Introduction} \label{introduction}

The rise of stablecoins represents a significant step in the merging of conventional finance and the digital economy. With a total market capitalization nearing \$130 billion, these digital assets have become pivotal for the dynamic interplay of economic forces across different markets: Processed stablecoin volumes are challenging major global payments incumbents (Carter, 2022) \cite{carter2022}, such as Visa or PayPal. Stablecoins also act as conduits for the transmission of interest rates into the digital economy. For example, the platform Flux Finance, which facilitates permissionless lending of USDC secured by tokenized U.S. Treasuries, transmits interest rates through stablecoin users, who lend or borrow against tokenized representations of U.S. treasury instruments. Similarly, MakerDAO's DAI Savings Rate Module,\footnote{Dai is an overcollateralized, decentralized stablecoin pegged to the USD.} bolstered by income from U.S. 
Treasuries, offers DAI holders an attractive store-of-value compared to other stablecoins.

However, stablecoins can also cause reverse transmission, with the potential of channeling impacts from on-chain markets to traditional off-chain markets. As substantial holders of U.S. Treasuries in their aggregate —surpassing\footnote{See \href{https://unchainedcrypto.com/tether-holds-more-us-treasury-bills-than-mexico-australia-spain-and-uae/}{here}.} that of entire nations like Australia— stablecoin issuers may adversely impact off-chain markets. For example, a liquidity  run on such a major stablecoin issuer could theoretically result in destabilizing markets if large amounts of U.S. treasuries were automatically liquidated in a short amount of time to meet redemption obligations.

The growing size of the stablecoin market, its importance for economic transactions in the digital and real world, as well as its dual role as a shock transmitter has not gone unnoticed by regulators, which are mandated with micro- and macroprudential regulation. Recognizing the importance of stablecoins, regulatory bodies around the globe are developing prudential frameworks to bring this emerging pillar of the financial system into scope.  

When developing a risk management framework for a stablecoin issuers, it is important to note that their business model closely resembles that of banks because they generate  profits through a net interest rate margin derived from maturity mismatches between ``deposits" and investments. Crucially, this overlap in business model and economic activity also exposes them to similar risks, such as duration, credit, counterparty, market, liquidity, and operational risks. However, though similar in operation —performing maturity and liquidity transformation— key differences remain. Unlike banks, stablecoin issuers are not in the business (yet) of issuing loans to support real economic activity.  They are also not covered by deposit insurance schemes.

Micro- and macroprudential regulation set by the Basel Committee on Banking Supervision (BCBS) aims to address such risks to financial stability, the economy and consumers by requiring banks by law to maintain capital buffers and follow specific risk management guidelines such as minimum liquidity buffers, as well as applying requirements on governance and compliance. Conversely, stablecoin issuers currently navigate these risks mostly outside the regulatory perimeter. That situation is addressed with initiatives like the EU's MiCA regulation and a proposed U.S. Stablecoin Bill, reflecting a legislative shift towards oversight that aims to safeguard stability and consumer protection within this evolving sector. 

To develop and implement a risk management framework for stablecoin issuers, two points are crucial. First, similar activity and risk exposure in the banking sector suggest that similar risk metrics as in established banking regulation may be applied. Second, public blockchain architecture, which is the infrastructure stablecoins operate on, provides key advantages for risk management frameworks, in particular regarding data quality, transparency and immutability, as well as scalable automation. 

In this paper, we develop a simple risk management framework for on-chain stablecoin protocols tailored to the DAI, the biggest decentralized stablecoin with a market capitalization exceeding \$5bn at the time of writing. The paper is organized as follows. Section \ref{literaturereview} provides a literature review. Section \ref{riskmetrics} outlines our proposed risk metrics. Section \ref{casestudy} shows how these can be conceptually used to monitor and manage risk in real time for MakerDAO. Section \ref{conclusion} concludes.

\section{Literature Review} \label{literaturereview}

The existing literature on stablecoins identifies risks and potential of stablecoins from several angles, such as systemic risk and consumer protection, and generally calls for stablecoins to be brought into regulatory scope.

Aldasoro, Mehrling and Neilson (2023) \cite{aldasoro2023} provide an analysis of stablecoins from the perspective of monetary economics, drawing parallels with traditional financial systems and emphasizing the need for similar safeguards in the digital asset space. Their paper reveals the basic nature of the on-chain liquidity mechanisms that underpin stablecoin models' promise of par settlement, identifying liquidity, rather than solvency, as the main concern for maintaining par value. The paper stresses the importance of regulatory oversight, including liquidity, solvency, and stability mechanisms to ensure stablecoins' ability to maintain parity with their underlying assets.

Catalini and de Gortari (2021) \cite{catalini2021} outline the economic design of stablecoins, emphasizing risk management and regulation. They discuss methods to address two crucial risks of stablecoin design: the volatility of reserve assets and the risk of a ``death spiral." The authors highlight the significance of backing stablecoins with high-quality liquid assets to mitigate volatility and argue for regulatory frameworks to protect against risks and ensure stability.

The Digital Euro Association (2022) \cite{digitaleuro2022} investigates the role of Euro stablecoins within the digital money landscape, focusing on their design, interaction with CBDCs, and the regulatory framework necessary for their integration into the eurozone's financial system. The findings include recommendations aimed at policymakers involved in EU crypto and stablecoin regulation, addressing the importance of legal and regulatory support for the introduction and management of stablecoins. The paper recommends ensuring transparency of stablecoin reserves, suggesting that issuers provide auditable evidence that their collateral is liquid and fully covers issued stablecoins.

The Financial Stability Board (2019) \cite{fsb2019} presents regulatory issues regarding stablecoins, emphasizing the importance of monitoring crypto-asset developments and managing potential risks. It highlights that stablecoins can become of systemic importance, combining characteristics of various financial services and potentially giving rise to new stability risks. Key concerns include the soundness of stablecoin reserves, their impact on traditional banking mechanisms, and the potential for adverse confidence effects due to issues like market manipulation and money laundering. The report acknowledges the benefits of stablecoins, such as reducing transaction costs and facilitating financial inclusion, but emphasizes the necessity of comprehensive regulation and oversight. It calls for a clear understanding of stablecoin arrangements and their legal implications.

Liao and Caramichael (2022) \cite{liao2022} discuss the growth of stablecoins and their potential to become a breakthrough in payment innovations, analyzing their impact on the banking system. They argue that dollar-pegged stablecoins, backed by safe and liquid collateral, exhibit safe asset qualities, particularly during episodes of market distress, but also underscore the risk of runs on stablecoins backed by non-cash-equivalent risky assets. The paper also discusses the impact of stablecoin adoption on traditional banking, varying with the sources of inflow and the composition of stablecoin reserves. It explores several scenarios, including a two-tiered banking system supporting stablecoin issuance and a narrow bank approach, which, while providing a stable peg to fiat currencies, could lead to banking disintermediation. The paper emphasizes the importance of stablecoin reserves being adequately safe and liquid to serve as a digital safe haven and mitigate run risks. 

Ostercamp (2022) \cite{ostercamp2022} provides a comprehensive analysis of the regulatory landscape for stablecoins across the EU, UK, and U.S. He highlights the potential risks stablecoins pose to financial stability, market integrity, and consumer protection, emphasizing the need for effective regulation to mitigate these risks. The paper also discusses how existing financial regulations could be adapted to stablecoins, considering their unique technological characteristics. It proposes that stablecoin issuers should be subject to strict requirements on the investment and management of reserve assets, in addition to maintaining liquidity and capital buffers. This approach would ensure redemptions can always be met, negating the need for bank-like deposit protection.

Schwarcz (2022) \cite{schwarcz2022} outlines a legal framework for regulating global stablecoins. His paper emphasizes the necessity of a legal strategy to address the challenges posed by these digital assets, particularly in terms of reserve management and investment. It also outlines key regulatory goals, including consumer protection, privacy, and safeguarding monetary integrity and financial stability.

While the above overview of stablecoin research generally calls for regulation of stablecoins, it does not provide specific implementation proposals, neither regarding risk metrics nor how advantages of public blockchains, such as neutrality, programmability and high quality, immutable real-time availability of data, can be harnessed for regulatory purposes. The two following papers, which also argue for the regulation of stablecoins, provide more insights in this respect and are therefore closer related to our following analyses.

First, Arner, Auer and Frost (2020) \cite{arner2020} examine the risks and potential applications of stablecoins, especially in the context of decentralized systems. They suggest that regulatory approaches to global stablecoins should consider their diverse uses and the integration of stablecoins as monetary instruments in digital environments. Key findings include the need for stringent regulations to address financial stability and manage conflicts of interest, particularly for global stablecoins which pose higher risks. The paper underscores concerns about market integrity, investor protection, the necessity of asset segregation and collateral security. It also highlights the importance of cross-border regulatory coordination due to the digital and borderless nature of stablecoins. Specific regulatory treatment is recommended for systemically important stablecoins, with attention to operational and cybersecurity risks. 

Additionally, the paper discusses the potential impact of stablecoins on the traditional banking sector, including the risk of disintermediation and increased costs in traditional bank lending. The concept of ``embedded supervision" is introduced, utilizing public blockchain ledgers for compliance monitoring, which could enhance regulatory oversight while reducing the reporting burden. The framework we develop in this paper can be seen as a first step to implement embedded supervision, which informs the regulator in real time of microprudential metrics of individual stablecoin protocols. 

Second, Bertsch (2023) \cite{bertsch2023} identifies the inherent vulnerabilities of stablecoins, particularly the risk of runs driven by concerns about reserve quality, liquidity, operational, technological, and custodial risks. The paper also discusses a theoretical model exploring the fragility of stablecoins, indicating that adoption can affect fragility by changing the composition of stablecoin holders, eventually making the system more susceptible to runs. It also notes that factors like the liquidity of reserves, bankruptcy costs, and seigniorage income influence this fragility. 

The author suggests that a wider adoption of stablecoins could lead to concerns about financial stability and the disintermediation of banks. To manage these risks, he advocates for additional regulation, including capital requirements and measures to improve reserve quality, to mitigate operational and technological risks. In particular, the author recommends that issuers should hold 100\% reserves in high-quality liquid assets against their redemption liabilities. Additionally, an extra capital cushion is advised to cover operational losses, asset price declines, or runs on the stablecoin. Our work proposes two specific risk metrics, for capital and liquidity, which have also been identified by Bertsch as crucial for microprudential regulation of stablecoin protocols.

Our paper also contributes to the literature by providing a simple risk and balance sheet management framework and associated risk appetite statement (including corresponding limits) for stablecoin platforms.

\section{CALM Risk Metrics and Framework} \label{riskmetrics}

Building on the insights from the literature review, which highlight liquidity and capital/solvency as key axes for stablecoin regulation, we propose a risk management framework for on-chain stablecoin protocols. This system is inspired by the Basel regulations' emphasis on liquidity and capital and takes full advantage of the blockchain’s liveness, tamper resistance, and transparency—qualities that are intrinsic to the technology and beneficial for prudential outcomes.

To fully benefit from these advantages, it is crucial to note that our focus here is on on-chain operated stablecoin protocols -as opposed to stablecoin issuers, such as Circle or Tether, who essentially tokenize fiat currencies. In particular, since the latter hold most of their reserves off-chain, a large part of their balance sheet (their assets) is obscure and cannot benefit from on-chain data advantages in a straightforward way (it would not be impossible though to use oracle technology as a second best solution).\footnote{Further differences pertain to underlying trust assumptions. In the case of a fiat-backed stablecoin, the minting process is operated in a “trusted” way. In other words, at some crucial point of the minting process, the user has to trust the issuing entity to complete the process. Typically, the process cannot be completed without some form of custodial reconciliation or centralized control. In the case of a decentralized stablecoin, the minting and redemption processes are handled in a so-called ``trustless'' manner by the end-user, provided the smart-contracts that mediate these interactions are open, neutral and public. Users are exposed to technical risk, but have a different, possibly mitigated, level of counterparty risk.}

Our proposed framework sets out to establish a simple yet robust mechanism for real-time tracking of risk metrics. Hopefully this framework can contribute to a growing body of regulatory research with traditional financial oversight metrics enhanced through the possibilities of public blockchain technology. In the following we introduce two specific metrics, capital at risk and the liquidity coverage ratio.

\subsection{Capitalization} \label{capitalization}

To assess a stablecoin’s underlying capitalization, we make use of the concept of capital at risk (CaR), which estimates potential losses based on i) historical data or hypothetical scenarios, and ii) the present balance sheet. 

As shown in Equation \ref{eq1}, the CaR is defined as the sum of the product of each asset exposure and its corresponding Capital at Risk Ratio (CaRR). The CaRR is defined as the sum of specific risk ratios, namely duration, credit, (crypto) market and operational risk.

\begin{equation} \label{eq1}
CaR=\sum^{n}_{i}{Exposure(i)} \cdot CaRR(i)
\end{equation}

\begin{equation} \label{eq2}
CaRR(i)=Duration(i)+Credit(i)+Market(i)+Operational(i)
\end{equation}

For duration risk, we simulate the impact of interest rate shocks on capital. This is similar to the approach taken in Basel III SRP 31.90.\footnote{See \href{https://www.bis.org/basel_framework/chapter/SRP/31.htm\#paragraph_SRP_31_20191215_31_90}{here}.}

For credit risk, we explore two methods. One is based on loss calculations (probability of default (PD) times loss given default (LGD)), which allows for more degrees of freedom in modeling individual credit risks. The other is a simplified approach using risk ratings to assign credit risk (similar to CRE20\footnote{See \href{https://www.bis.org/basel_framework/chapter/CRE/20.htm?inforce=20230101&published=20221208}{here}.} in Basel III). 

For (crypto) market risk, used for crypto-backed loans, the parameter is derived from the work of Blockanalitica, which allows estimating capital losses based on Monte Carlo simulations of (crypto collateral) price drops.\footnote{Crypto-backed loans consist of locking crypto-collateral in a smart contract to borrow a stablecoin. Because of pseudonymity –borrowers can usually not be identified– the collateral  must exceed the amount borrowed (“over-collateralisation”). For example, if a user wants to borrow DAI worth \$100, she can lock Ether worth \$200 into a MakerDAO smart contract. If the collateral value falls below a specified threshold relative to the borrowed value, the collateral is liquidated, i.e. sold on the market to repay the loan. Under extreme market fluctuations, it is possible that the collateral is not sold fast enough to cover all the loan - in that situation the protocol is exposed to bad debt and loses capital. See Section \ref{casestudy} for further details on how this process works in MakerDAO as an example}  

For operational risk, we use a qualitative assessment of the risks involved, which can be manifold. For example, smart contracts have unique technical risks that can expose stablecoins to operational risks. Complex off-chain processes, which need custodians and other service providers may add another layer of operational risk.

Following equations \ref{eq1} and \ref{eq2} above, $CaR$ is the aggregate expected loss of capital from the four possible risks outlined above. For example, if $CaR=100$, then the expected loss of capital is $100$. Based on that metric, we define the capitalisation ratio, $CR$, as follows: 

\begin{equation} \label{eq3}
    CR=\frac{Capital}{CaR}
\end{equation}

where $Capital$ is the current capital of the stablecoin protocol. As long as $CR\geq 100\%$, the expected capital loss does not exceed the stablecoin's capital buffer.

Based on our metric, there are two outcomes:  
\begin{itemize}
\item If $CR \leq 100\%$ the protocol is considered undercapitalized,
\item If $CR > 100\%$ the protocol is considered sufficiently capitalized.
\end{itemize}

The target level of the CR is discretionary. While it would not make sense to set it under 100\% –objectively a solvency limit– a stablecoin could conceivably operate under more conservative parameters and include a buffer above the minimum required capital level, such as requiring a $CR > 105\%$.

\subsection{Liquidity Coverage Ratio} \label{liquidity}

The ‘socket theory’ (‘Bodensatztheorie’) provides a historic but intuitive rationale for banks’ maturity transformation and liquidity management, which can be used to motivate a simple metric for liquidity risk management of stablecoins. Following that theory, (Wagner, 1857) \cite{wagner1857} points out that while the ``contractual'' maturity of demand deposits is daily, many customers leave their deposits for much longer periods without withdrawing them. At the same time, some deposit outflows, which actually take place, are substituted by inflows of demand deposits. Since a bank has a large number of customers and deposits, in-and-outflows are independent, the law of large numbers can be used to predict a stable base of deposits, which the bank can use for longer term investments, that is, carry out maturity transformation. Existing liquidity regulations ultimately stem from the considerations laid out in that basic theory. For example, (Choudhry, 2007) \cite{choudhry2007} proposes, among other approaches to:

``divide deposits into stable and unstable balances, of which the core deposits, which are determined by the bank based on a study of the total balance volatility pattern over time, are set as a ‘permanent’ balance with longer maturity. The excess over the core balance is then viewed as very short-term debt.''

More advanced Basel III regulations, such as the Liquidity Coverage Ratio (LCR), follow a more sophisticated modeling approach and measure a financial institution's ability to handle short-term liquidity demands. It is defined as the proportion of high-quality liquid assets (HQLA) to be held to cover projected cash outflows over a 30-day stress period.

When developing a simple liquidity risk metric for stablecoin protocols, some caveats need to be kept in mind.  First, pseudonymity makes clustering liabilities (the issued stablecoin) less reliable than possible for banks, who may know from their customers’ identities if several deposit accounts belong to the same person or entity.  Second, as stablecoins are uniquely composable with other smart contract interactions, customers can send their stablecoins to other products, such as an automated market-maker (AMM) exchange, and still maintain custody. Finally, decentralized finance applications (and their users) are still quite young and user behavior data might change rapidly. Constructing an elaborate model based on a short time series of past behavior may lead to misplaced confidence.

It therefore makes sense to start with a simple yet robust modeling approach based on the above principles. Crypto-issued stablecoins have the advantage that HQLA needs can be estimated from user behavior simply by observing on-chain transactions. We divide stablecoin holdings into two buckets, ‘volatile’ and ‘organic’ (or the ‘socket’ in (Wagner, 1857)). Volatile refers to stablecoins held in smart-contract wallet addresses that interact with applications, such as lending or trading, which in the past have indicated more ‘flighty’ customer behavior in the event of a market crash. By contrast, organic stablecoins refer to those held in externally-owned addresses (EOA), possibly indicating abstention from speculative activity.

Following this approach,\footnote{An example can be consulted  \href{https://forum.makerdao.com/t/modeling-dai-maturity/}{here}.} the maximum historical drawdown of stablecoins in each holding type (volatile and organic) can be measured based on historical data. Periods of higher volatility, such as during May 2022, provide a useful backdrop for stress-testing the maximum likelihood that a category of wallet address will draw down, i.e. redeem or sell, its stablecoin holdings under relatively large shocks. These holding patterns can be further classified into multiple time buckets for the organic part. We propose to use the following buckets: up to 1-day, up to 1-week, up to 1-month, and up to 1 year.

Bucketing stablecoin issuers’ liabilities as outlined above provides insights about the rough maturity structure and can help inform about the required amount of liquidity to fulfill redemption demand under stressed conditions. In particular, we introduce the funding gap as a simple metric to measure liquidity risk over time bucket $t$:

\begin{equation} \label{eq4}
    FG_t=Outflow_t-Liquidity_t
\end{equation}

where ``Outflow" is the maxiumum expected outflow under stressed conditions and ``Liquidity" the available funds to satisfy the drawdown, that is, liquidity available to maintain the peg. 

Liquidity in bucket $t$ can be computed by approximating the amount of fiat-backed stablecoins, which can realistically be obtained with the available assets under stressed conditions and depends on asset liquidity and the sales process. For example, we consider that USDC\footnote{USDC is a centralized, fiat-backed stablecoin issued by Circle Internet Financial LLC. Its reserves are held entirely off-chain in bank deposits and short-term US treasuries. Conversely, Dai is, by design, a decentralized overcollateralized stablecoins. Its reserves are created by users depositing crypto assets, including other stablecoins, into MakerDAO.} has the highest possible liquidity –given the existence of a liquid primary market– while large amounts of longer duration assets may be subject to slippage in the process of liquidation under stressed market conditions. A negative funding gap in bucket $t$ indicates that redemption demand under stressed conditions may exceed available liquidity to satisfy it, potentially putting downward pressure on the peg.  The notable difference with traditional bank management practice is that the tenor of the liabilities can be derived much more precisely (relative to bank deposits) from on-chain user behavior, and the composition of the assets can also be tracked with public data. This makes estimating the funding gap an exercise accessible to anyone.

\section{Case Study: A CALM Approach to Real-Time Risk Metrics for MakerDAO} \label{casestudy}
Evaluating the robustness of a stablecoin issuer using our proposed simple risk metrics involves testing its resilience against various historical and/or hypothetical risk scenarios. In this section, we provide the specific example of MakerDAO, the biggest on-chain stablecoin protocol, to show how our risk metrics can be applied to implement transparent monitoring of capital and liquidity risks in real time.
\subsection{Overview of MakerDAO} \label{overviewmaker}

MakerDAO is a Decentralized Autonomous Organization (DAO) on the Ethereum blockchain, designed to enable the decentralized creation and management of DAI, a stablecoin pegged to the U.S. dollar. It has a market capitalization of \$5.3bn at the time of writing. DAI maintains its value close to one U.S. dollar through a system of smart contracts. These allow users to mint DAI in two ways. 

Firstly, by depositing crypto collateral in excess of the amount of DAI they wish to generate and drawing an automatic loan at a given collateralization ratio. For example, a user can lock ETH worth \$2000 in a smart contract vault to obtain DAI1000. In the example, the ‘loan’ is overcollateralized by 100\%. Note that the user’s collateral would be liquidated if the collateral ratio drops below a specific collateral ratio, say 170\%.\footnote{Vaults have different collateral ratio requirements, depending on the type of collateral and other parameters.} Secondly, by directly exchanging USDC and into DAI via a so-called Peg Stability Module (PSM), which guarantees a 1:1 trade where anyone with the necessary technical and financial capabilities can engage in this process based on financial motivation, leveraging traditional market dynamics through arbitrage bots. 

This system ensures that DAI remains backed by assets that considerably exceed the issued DAI in value, aiming to preserve stability and trust in its proposition as a digital currency. MakerDAO's ecosystem includes mechanisms for parameter governance by its token holders, and automated risk management processes to safeguard the users of the protocol and the peg of DAI to the reference asset (USD).

Two mechanisms are key to maintain the peg to the USD. The first is a passive mechanism that requires users to lock up collateral as described above, typically Ethereum, in excess of the amount of DAI they want to generate — a process known as over-collateralization. While they ‘borrow’ DAI, a borrowing fee, fixed by governance, will accrue. This system ensures that when DAI's price deviates from its intended \$1 peg, market dynamics are triggered to correct the imbalance. For instance, a price drop below \$1 encourages users to purchase DAI at a cheaper rate on the market to settle their debts, which decreases the supply of DAI and helps elevate its price back to the peg. In contrast, when the price climbs above \$1, users are incentivized to increase their debt by minting more DAI to sell against another stablecoin, e.g. USDC, in the expectation of a return to peg to realize a profit. This increases its supply and helps to return to peg.  

Second, the DAI Savings Rate (DSR) serves as an indirect lever to modulate the circulating supply of DAI. By depositing DAI into a savings contract and earning rewards from MakerDAO's surplus, aggregate demand and supply dynamics for DAI can be influenced. A higher savings rate increases demand and therefore its price, incentivizing its creation, while a lower rate does the opposite: having a lower return, users will tend to sell DAI on the market, decreasing its price. This rate adjustment allows for a gradual regulation of DAI's supply and demand, contributing to the overall stability of its value.

The two mentioned parameters, borrowing and saving rate were the only present at the origin of MakerDAO. They allowed DAI to remain pegged, even if not perfectly, to the U.S. Dollar. With the increasing demand for DAI by the market over time, an additional mechanism to stabilize the peg was introduced: the PSM mentioned above. It is an active tool that allows for direct stabilization of DAI's value. Through the PSM, users can exchange DAI with other stablecoins at \$1, managing its supply and price. 

When the price of DAI is high, the module facilitates the sale of DAI for other stablecoins, boosting its supply and thereby lowers its price. Conversely, when DAI is below the peg, users can buy it on the secondary market and sell it against another stablecoin via the PSM at a price equal to 1. This reduces its supply and helps its price recover. Note that the PSM has finite reserves, that is, this stabilization mechanism is conditional on sufficient funds in the PSM. Also note that the amount of stablecoins in the PSM effectively fulfills the role of a HQLA. 

\begin{figure}[h]
\centering
\includegraphics[width=\textwidth]{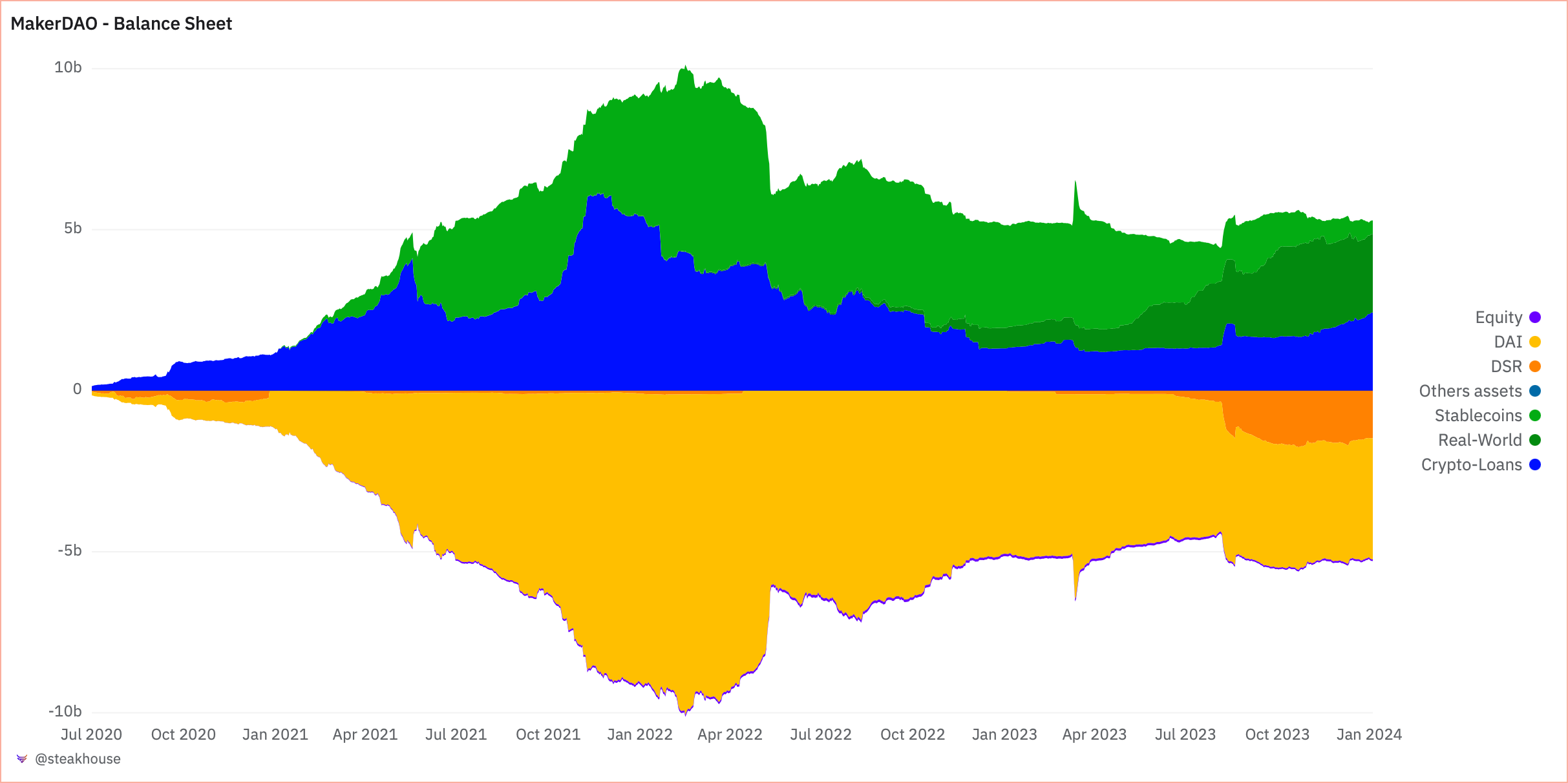}
\caption{MakerDAO’s Balance Sheet over time (see \href{https://dune.com/queries/2840463/4743803}{[Dune]})}
\label{fig:balance}
\end{figure}

Figure \ref{fig:balance} displays MakerDAO’s real-time balance sheet\footnote{The figure, which is available in real time \href{https://dune.com/queries/2840463/4743803}{here}, is based on an open querying tool called Dune Analytics, that allows users to write SQL queries to interact with historical publicly available blockchain data. MakerDAO’s smart contract function executions are recorded, aggregated and classified to produce the graph.} over time. Assets are displayed above the abscissa and liabilities below. For example, on 15 December 2023, liabilities consisted of ~\$3.6b DAI, ~\$1.6b in the DSR, and ~\$54m in equity. Assets consisted of ~\$2.6b of real-world assets (for example public debt), ~\$2.2b crypto-backed loans, ~\$0.54b of stablecoins, and ~\$49000 of other assets. Between July 2020 to January 2022, the balance sheet expanded (to about \$10b), and since then contracted to roughly \$5b at the time of writing. 

The open and real-time nature of decentralized stablecoins protocols means that the simple risk metrics, which we introduced in the previous section, can be tracked in real-time for MakerDAO. Steakhouse Financial is an example of an ``Ecosystem Actor'', or an independent business approved and engaged by DAO token holders to conduct these types of analysis and issue recommendations for parameter changes.

At the moment, a large proportion of MakerDAO's assets are held in off-chain U.S. treasuries. Information on outstanding notional amount, interest and payment flows is brought on-chain in a ``trusted'' setup to inform the risk framework.\footnote{See \href{https://dune.com/steakhouse/makerdao-clydesdale}{here} for an example.} For instance, as part of Steakhouse Financial's mandate to provide accessible and updated information, the DAO authorized it to receive read-only access to off-chain accounts to collect data, which are then regularly published on Dune Analytics. In the future, a clear improvement, in particular regarding the potential need for off-chain liquidations, would be for MakerDAO to rely on tokenized instruments with liquid on-chain markets instead.\footnote{See \href{https://forum.makerdao.com/t/next-steps-tokenized-t-bills/}{here for an example request for proposals.}}

\subsection{CALM Risk Metrics for MakerDao} \label{riskmetricsmaker}
In this subsection, we tailor our two risk metrics for MakerDAO to carry out a protocol risk assessment. We first calculate the CaR following Equations \ref{eq1} and \ref{eq2} based on the following risk parameterizations:
\begin{itemize}
    \item Duration risk is modeled as the expected capital loss from a 200bps upward shock to all interest rate sensitive balance sheet positions. Note that given MakerDAO’s current balance sheet composition (liabilities have zero duration, while its assets have positive duration), this is a conservative scenario.
    \item Credit risk is modeled as a loss of credit positions, such as bonds. We use $1\%$ $CaRR$, defined as $\frac{CaR(i)}{Exposure(i)}$, for a EUR AAA covered bond used by Société Générale Forge, 5\% for mortgages issued by a regulated community bank in the U.S. and 10\% for other assets with credit risk. This simplified approach is similar to Basel's CRE20, though the specific numbers may differ to account for varying types of risk exposure that Maker DAO faces.  
    \item Crypto market risk is based on Monte Carlo simulations that estimate realistic collateral price decreases under adverse market conditions.\footnote{See \href{https://medium.com/block-analitica/maker-capital-at-risk-model-methodology-2423b0c5913c}{here} for an outline of the methodology used.}
    \item The operational risk parameter is set to 0, other than a nominal amount to reflect operational risk in Coinbase Custody exposure. 
\end{itemize}

An overview of balance sheet items with exposure and capital at risk (see Equation \ref{eq3}) is provided in Table \ref{tab:car1}. 
\begin{table}[h!]
    \centering
    \begin{tabular}{|l|r|r|r|}
        \hline
        \textbf{Assets} & \textbf{Exposure (\$M)} & \textbf{CaR (\$M)} & \textbf{CaRR} \\
        \hline
        Crypto-Backed Loans & 2,380 & 68.2 & 2.9\% \\
        Public Credit & 2,317 & 11.9 & 0.6\% \\
        Stablecoins & 260 & 2.6 & 1.0\% \\
        Private Credit & 263 & 44.7 & 17.0\% \\
        \hline
        \textbf{Balance Sheet} & \textbf{5,221} & \textbf{128.9} & \textbf{2.44\%} \\
        \hline
    \end{tabular}
    \caption{Capital at Risk for MakerDAO on 31 December 2023 (see \href{https://dune.com/queries/3327701}{[Dune]})}
    \label{tab:car1}
\end{table}

Risk sources are not evenly distributed across asset types. Crypto-backed loans are, in absolute terms, the main source of risk with \$68.2M However, private credit is a relatively riskier component on the balance sheet, featuring a $CaRR$ of 17\%. Nevertheless, due to a relatively small exposure (\$263M), the $CaR$ is comparatively smaller (\$44.7M). Private credit features the highest $CaRR$ because it has significant duration risk (due to a longer average maturity) and credit risk. On the other side of the spectrum, public credit (T-bills, with an average maturity of 3 months) only has relatively small duration risk. Lastly, fiat-backed stablecoins are assigned an arbitrary 1\% credit risk, primarily to reflect idiosyncratic credit risk in stablecoin issuers (mostly USDC and USDP), despite being largely backed by public debt and cash deposits. The last row in Table \ref{tab:car1} shows the aggregate CaR for the balance sheet.

Table \ref{tab:car2} breaks down the aggregate $CaRR$ (last column in Table \ref{tab:car1}) into the individual risk components. For example, summing up all four risk parameters for $CaR$ for private credit, one obtains 17\%.

\begin{table}[h!]
    \centering
    \begin{tabular}{|p{4cm}|p{2cm}|p{2cm}|p{2cm}|p{2.3cm}|}
        \hline
        \textbf{Assets} & \centering\textbf{Duration Risk} & \centering\textbf{Credit Risk} & \centering\textbf{Crypto Markets Risk} & \centering\textbf{Operational Risk} \tabularnewline
        \hline
        Crypto-Backed Loans & \centering 0\% & \centering 0\% & \centering 2.9\% & \centering 0\% \tabularnewline
        Public Credit & \centering 0.5\% & \centering 0\% & \centering 0\% & \centering 0\% \tabularnewline
        Stablecoins & \centering 0\% & \centering 1.0\% & \centering 0\% & \centering 0\% \tabularnewline
        Private Credit & \centering 8.9\% & \centering 8.1\% & \centering 0\% & \centering 0\% \tabularnewline
        \hline
    \end{tabular}
    \begin{flushleft}
        \footnotesize \ \ \ \  Note: A notional amount for operational risk has been assigned to stablecoins held in Coinbase custody.
    \end{flushleft}
    \caption{Capital at Risk Ratio by risk type contributors for MakerDAO on 31 December 2023}
    \label{tab:car2}
\end{table}

\begin{figure}[h!]
\centering
\includegraphics[width=\textwidth]{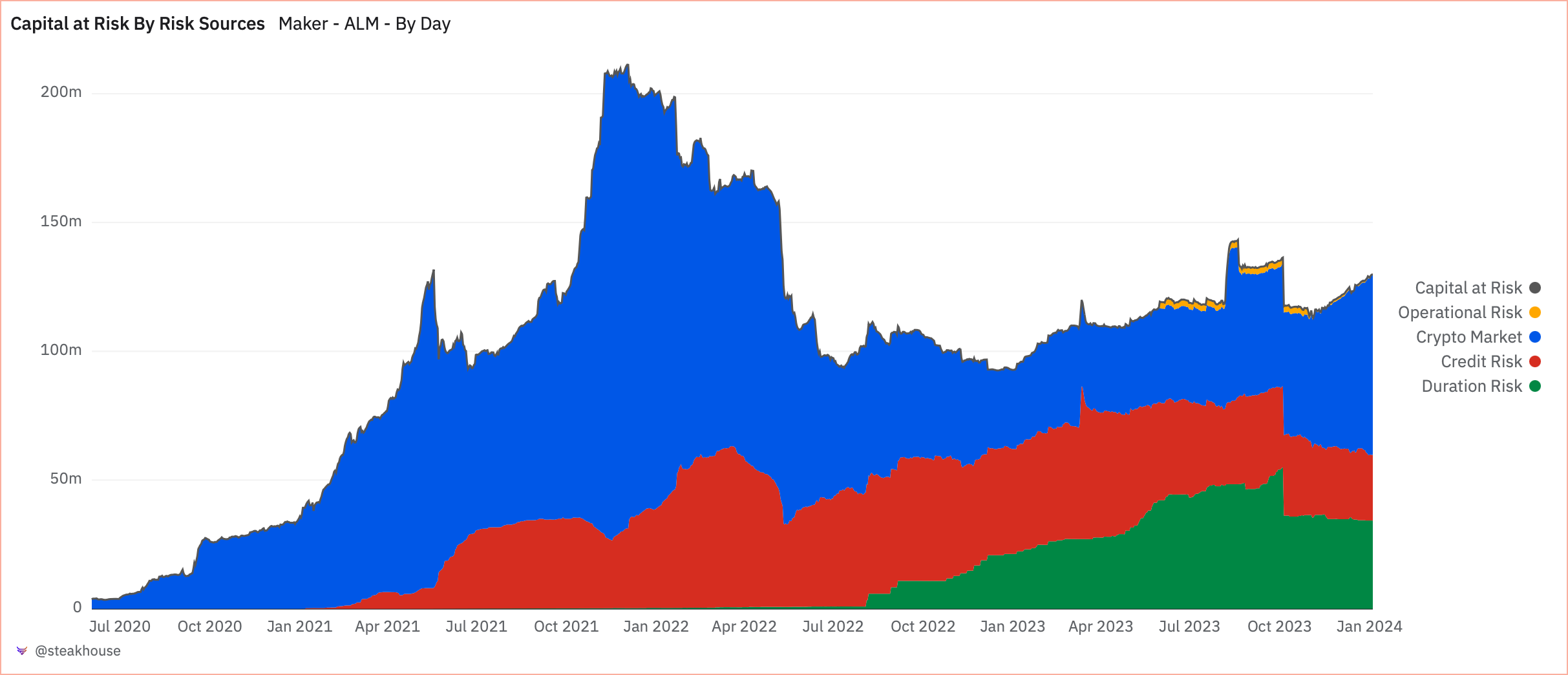}
\caption{Capital at Risk over time by risk source (see \href{https://dune.com/queries/3286476/5501452}{[Dune]})}
\label{fig:capital2}
\end{figure}

\begin{figure}[h]
\centering
\includegraphics[width=\textwidth]{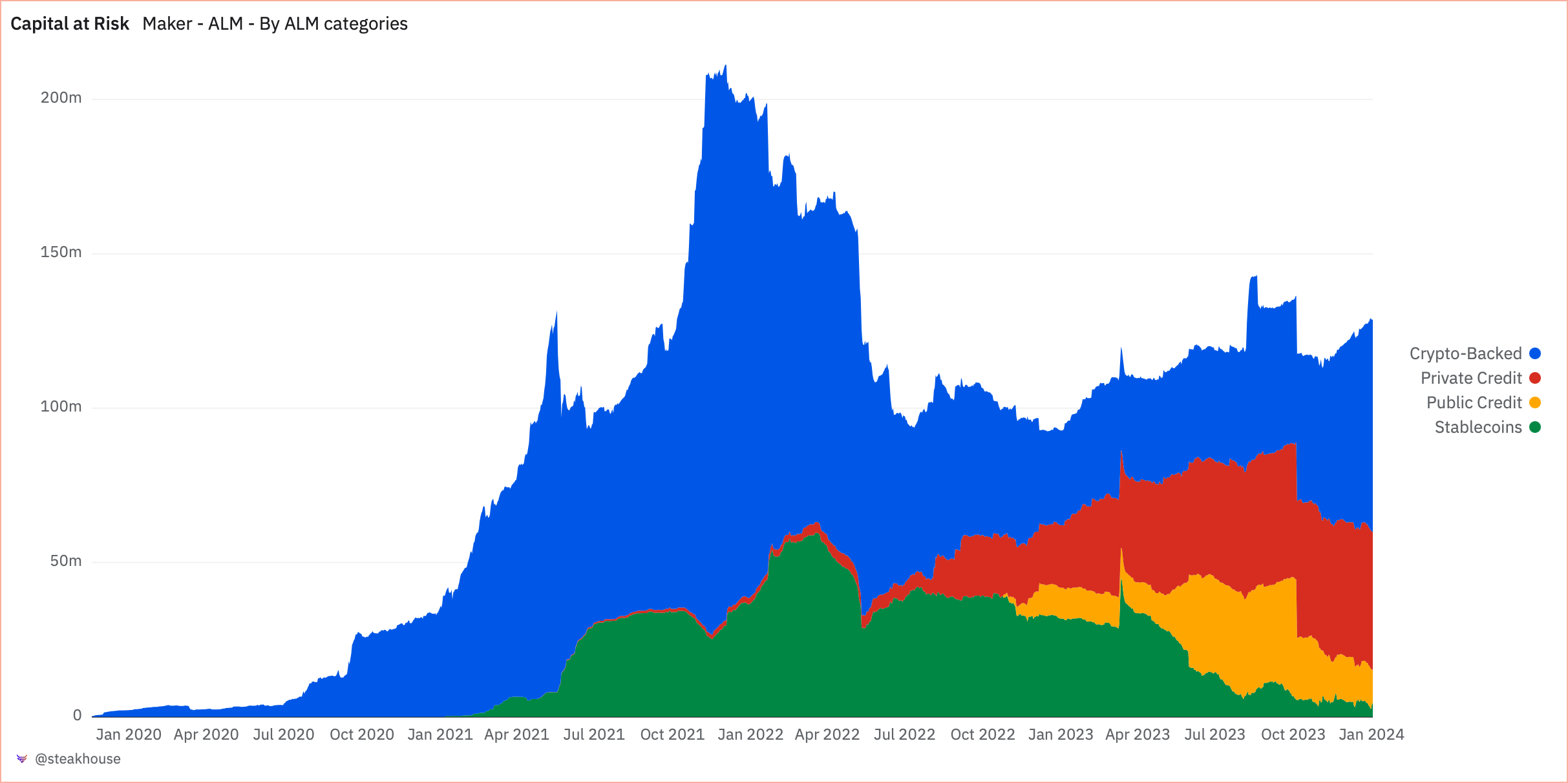}
\caption{Capital at Risk over time by asset type (see \href{https://dune.com/queries/3274044/5480144}{[Dune]})}
\label{fig:capital3}
\end{figure}

\clearpage

As displayed in Figures \ref{fig:capital2} and \ref{fig:capital3}, neither the balance sheet of MakerDAO nor the decomposition of risks\footnote{These charts are part of a comprehensive real-time dashboard (available \href{https://dune.com/steakhouse/makerdao-alm}{here}) provided by Steakhouse Financial and used by the MakerDAO community to monitor MakerDAO's capital at risk.} are static. Initially, all risk came from crypto-backed loans, which are exposed only to crypto markets risk. Subsequently, MakerDAO added so-called real-world assets, such as private credit, which at the time featured considerable credit risk (but initially relatively little duration risk). Starting at the end of 2022, MakerDAO allocated a growing portion of its balance sheet towards T-bills and short term treasury bond ETFs, thereby adding significant duration risk. In November 2023 the ETFs were sold and the duration was reduced to a 6-month T-bill ladder featuring an average duration of 3 months. Since then, some exposure to T-bills has been replaced by permissionless crypto-backed loans, which increased crypto-market risk. 

\begin{figure}[h!]
\centering
\includegraphics[width=\textwidth]{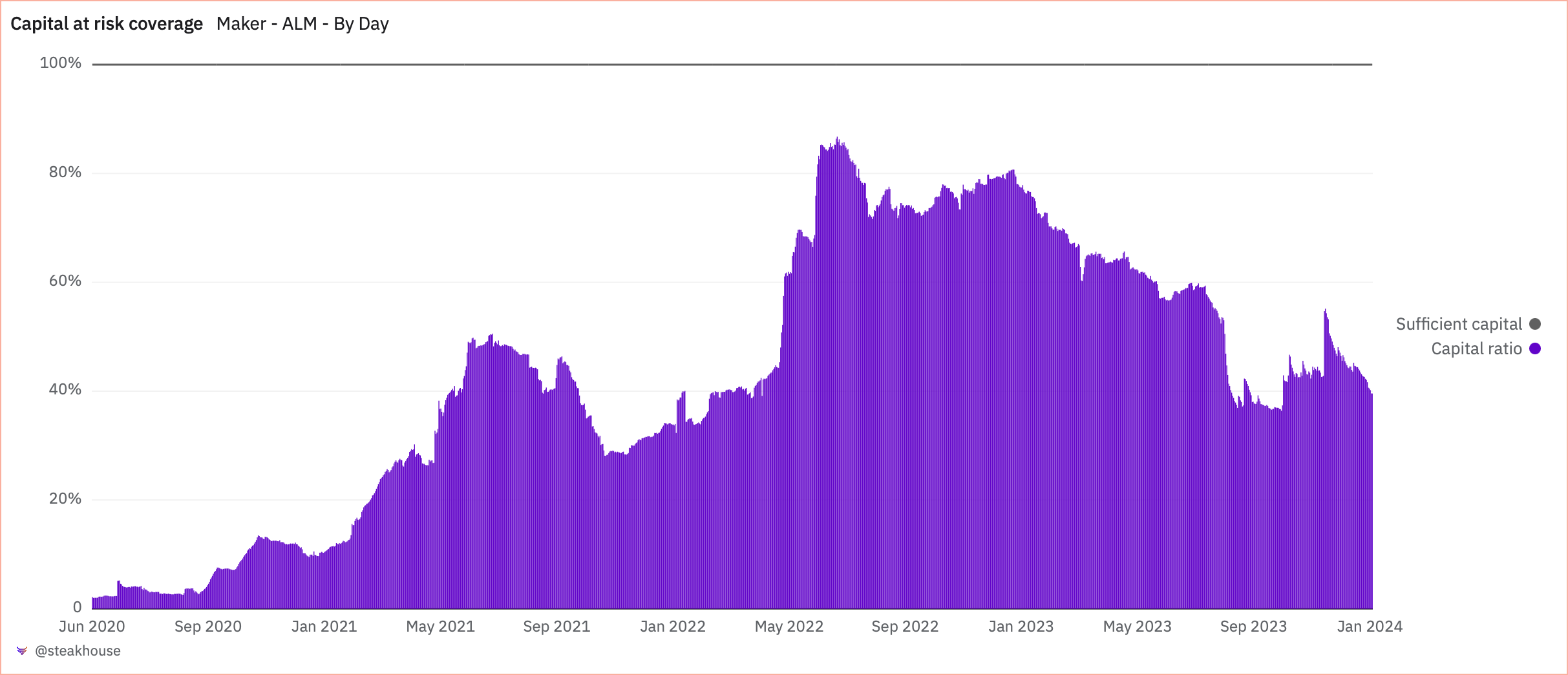}
\caption{Capital Ratio over time (see \href{https://dune.com/queries/3286476/5515381}{[Dune]})}
\label{fig:capital4}
\end{figure}

The capitalisation ratio (CR) (see Equation \ref{eq3}) over time is presented in Figure \ref{fig:capital4}. With a $CaR$ of \$128.7M for a baseline capital amount of only \$53.4M, the CR at the end of 2023 was at 41.6\%. That is, conditional on the adverse scenarios implied by the CaR, expected losses exceed the available capital. Note that MakerDAO started without capital and built a capital base over time. It is important to point out that our result neither means that i) MakerDAO is on the brink of default, nor ii) that the proposed shock scenarios covered in our metric are unrealistically adverse. MakerDAO has been resilient in the past to very large shocks, including those that resulted in the depegging of prominent stablecoin issuers. It therefore has stood the test of time in the arguably most adverse market that exists (and there are no bailouts in crypto). Our metric covers a multiple shock scenario, that effectively stresses all reserve positions with large shocks at the same time. While the probability of individually realistic large adverse shock scenarios realising all at the same time may be low, we think that it is prudent to err on the side of caution because user funds are exposed. We therefore propose the rather restrictive parameterization.

Depending on market sentiment, the demand for DAI and of crypto-backed loans increases or decreases, implicitly affecting the $CR$. The declining cryptocurrency market since the end of 2021, and with it a decrease in DAI issued due to lower demand, allowed the $CR$ to increase to almost 100\%. Nevertheless, the increased exposure to private credit, in combination with the return of a more favorable crypto environment (leading to an increase in DAI issuance via increasing demand for crypto-backed loans), decreased the capital ratio to its current level. At the moment, excess revenue generated by MakerDAO is used to reduce the amount of its governance token in circulation, akin to a share buyback with the aim to reward investors. Based on our findings, we recommend that the MakerDAO community explores mechanisms that channel income into the capital buffer as long as $CR < 100\%$. Note that when the $CR$ exceeds $100\%$, excess capital could be spent on buying the governance token on the secondary market to reward investors. 

\begin{figure}[h!]
\centering
\includegraphics[width=\textwidth]{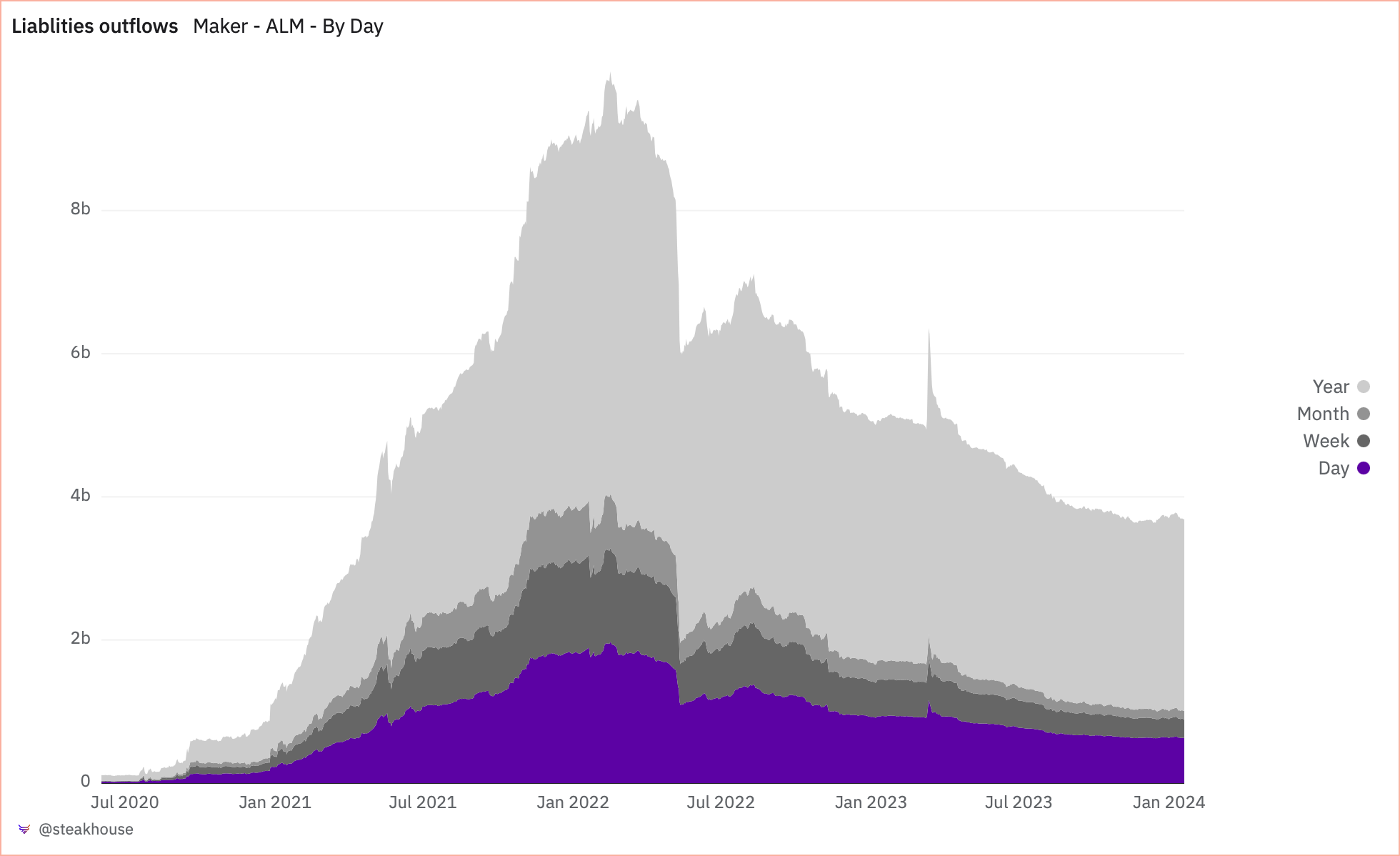}
\caption{Expected DAI redemptions under stress, by tenor (see \href{https://dune.com/queries/3286476/5515552}{[Dune]})}
\label{fig:capital5}
\end{figure}

\begin{figure}[h!]
\centering
\includegraphics[width=\textwidth]{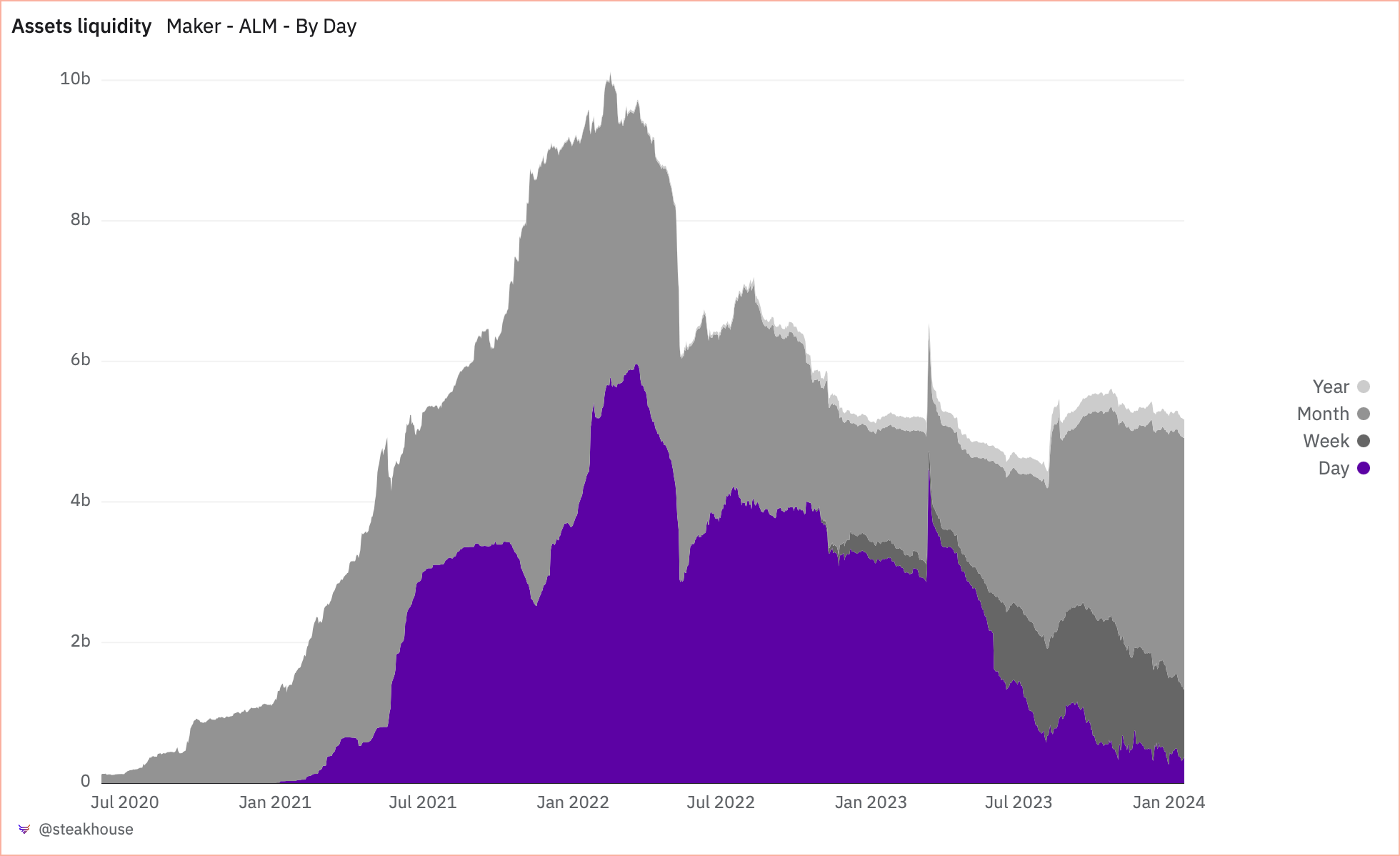}
\caption{MakerDAO Asset tenor evolution over time (see \href{https://dune.com/queries/3286476/5515565}{[Dune]}) on 31 December 2023}
\label{fig:capital6}
\end{figure}

\begin{figure}[h!]
\centering
\includegraphics[width=\textwidth]{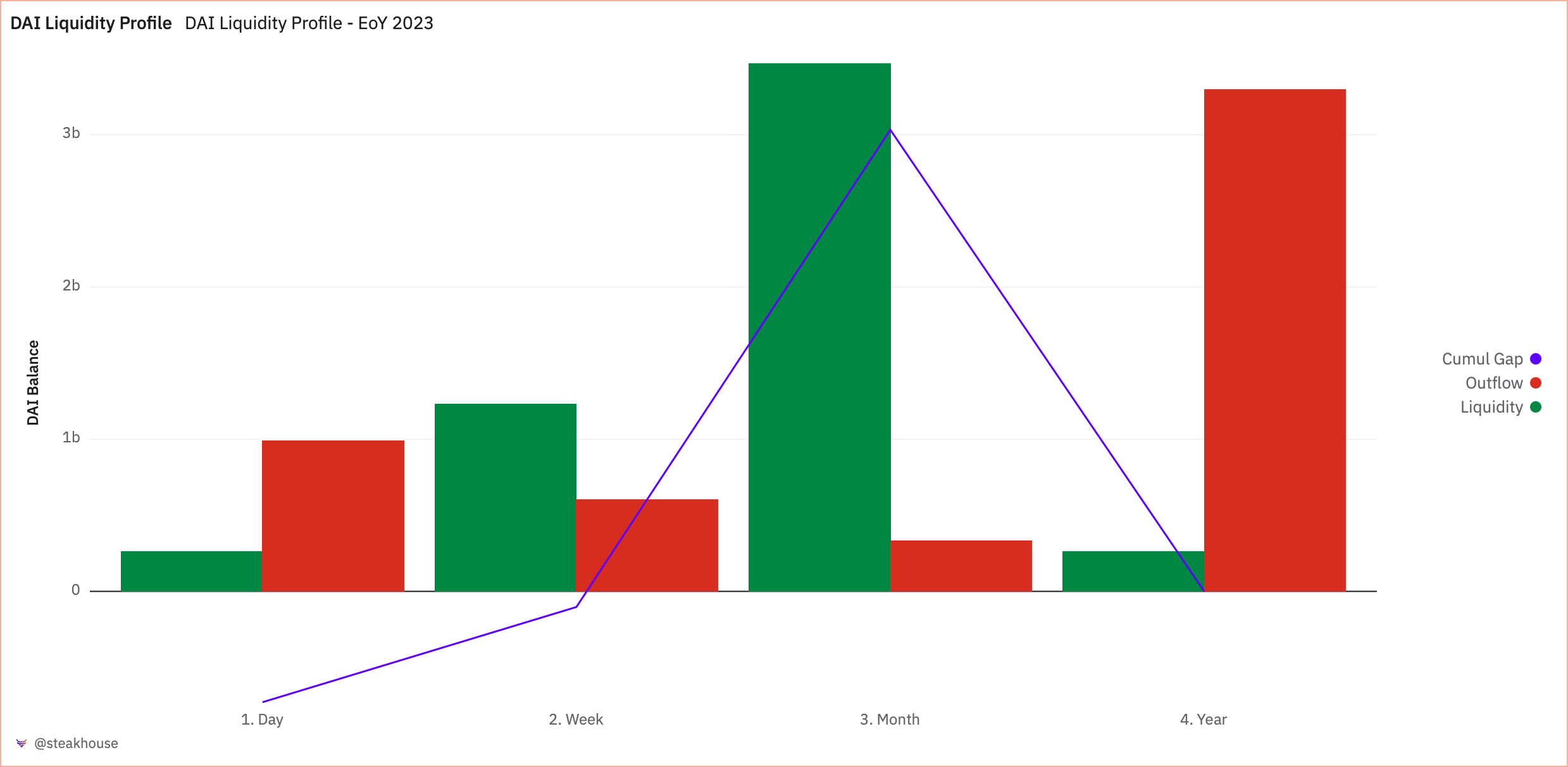}
\caption{Dai ALM funding gap chart (see \href{https://dune.com/queries/3209073/5365000}{[Dune]})}
\label{fig:gap}
\end{figure}

Next, we turn to our metric for liquidity risk, as displayed in Equation \ref{eq4}. Note that for simplicity, we do do not distinguish between volatile and organic liabilities in this analysis but slot all liabilities across our buckets. Figures \ref{fig:capital5} and \ref{fig:capital6} display the expected DAI outflows (``Possible Outflows") as well as available liquidity (``Available Liquidity"), respectively, both under stressed conditions across the four time buckets. For example, on 31 December 2023  the 1-day and 1-week buckets indicate a negative funding gap (-\$728M and -\$102M, respectively), which gets resolved starting at the 1-month  (~\$3,032M), as shown in Table \ref{tab:gap} and Figure \ref{fig:gap}. Note that the funding gap is cumulative, that is for example, the weekly funding gap includes the daily one. The funding gap ends up as null at the longest maturity bucket as all liabilities are matched with assets. This result indicates that MakerDAO has funding gap for the shortest buckets considered in our metric. Its peg could be under risk under stressed liquidity conditions. 

As can be gauged from Figures \ref{fig:capital5} and \ref{fig:capital6}, the asset tenor evolution appears relatively more volatile in comparison to the liabilities' tenors. The former is under control of the DAO, while the tenor of liabilities depends to a large extent also on user demand and preferences for DAI. These appear to shift rather gradually over time. Therefore, MakerDAO can exert control over the funding gap metric to reflect (gradual) changes in the maturity structure of liabilities. For example, if a persistent gap develops in the shortest bucket, funds in the PSM may be gradually increased (by re-allocating longer term assets) to close the funding gap. Based on this insight, we recommend that the MakerDAO community considers increasing the available liquidity in the daily bucket by $\$728M$.\footnote{Since the funding gap is cumulative, this would automatically result in a positive funding gap in the weekly bucket as well.} Conversely, the funding gap framework suggests that the asset allocation could afford to take more maturity exposure on the basis of much greater liquidity (ca. \$3bn) in the longer maturity buckets to improve capital returns.

\begin{table}[h!]
\centering
\begin{tabular}{|l|r|r|r|}
\hline
\textbf{DAI Maturity} & \textbf{Possible Outflows} & \textbf{Available Liquidity} & \textbf{Funding Gap} \\
\hline
1. Day            & 988,496,652      & 260,424,144        & \textcolor{Maroon}{-728,072,509} \\
2. Week           & 603,190,592      & 1,228,440,892      & \textcolor{Maroon}{-102,822,208} \\
3. Month          & 333,981,595      & 3,468,811,411      & \textcolor{ForestGreen}{3,032,007,607} \\
4. Year           & 3,295,302,537    & 263,294,929        & \textcolor{ForestGreen}{0} \\
\hline
\end{tabular}
\caption{MakerDAO Funding Gap Liquidity Metric (see \href{https://dune.com/queries/3209073/5364789}{[Dune]})}
\label{tab:gap}
\end{table}

\clearpage

\section{Conclusion} \label{conclusion}
This paper is a first step towards developing an (automated) on-chain risk management and crypto asset-liability management (CALM) framework for stablecoin protocols. 

Our two metrics proposed, capital at risk and funding gap, are built on the foundations of orthodox asset-management practices drawn from bank management and existing regulatory frameworks such as the Basel set of rules. This approach, which is possible in real-time using high-quality, granular data that come as a ``by-product" of public blockchain technology, allows for a paradigm shift in risk management. We demonstrate, using the example of MakerDAO, how risk monitoring and management can be carried out transparently, automated and in real-time for on-chain protocols.

Based on our metrics and findings, we provide two recommendations. Firstly, we recommend that the MakerDAO community considers exploring dynamic re-capitalization based on the $CR$ over time. For example, MakerDAO could add any profit to the capital buffer as long as $CR<1$, and, conversely, spend its capital buffer if $CR>1$. This would automatically result in adequately capitalising MakerDAO over time and effectively mimic Basel policies that prevent banks from paying out dividends, if their capital buffer is below a certain threshold. 

Second, based on our finding for liquidity risk, we recommend the MakerDAO community explores increasing fiat-backed stablecoin liquidity in the PSM by ~\$728M and conversely consider extending maturity exposure to close a positive long-term \$3bn funding gap as well. It may also be worth exploring a set of guiding principles that aim to match expected outflows under stress with available liquidity in respective time buckets over time.

Our findings may also be useful for regulators, who are currently exploring how to bring issuers of fiat-backed centralized stablecoin into scope. Requiring the use of similar metrics for monitoring and risk management purposes could benefit regulators in developing real-time monitoring tools for a new asset class that is set to considerably grow in the near future. In subsequent papers we could explore further refinements and automations that could be implemented at MakerDAO, as a way of continuing development of this asset-liability management framework. 

In a forward-looking scenario, if more financial institution activities were to transition to public blockchains, it might be feasible to apply similar principles to systemically important institutions. This shift could enhance collateral transparency and enable real-time prudential monitoring. Such advancements may not be a panacea, but they could potentially reduce the risk of incidents similar to those experienced by Silicon Valley Bank recently, contributing to a more resilient financial system.

\medskip
\newpage
\printbibliography
\end{document}